\documentclass[aps,prl,twocolumn,groupedaddress]{revtex4}

\usepackage{graphicx}
\usepackage{amssymb}
\usepackage{amsmath}
\usepackage{amsfonts}
\usepackage{amssymb}
\usepackage{graphicx}
\usepackage{hyperref}
\usepackage{color}

\newcommand{\rs}{\rm \scriptscriptstyle}

\begin{document}

\title{Three-body Interactions in One Dimension}

\pacs{}

\author{B. Capogrosso-Sansone$^{1}$, S. Wessel$^{2}$, H.P. B\"uchler$^{2}$, P. Zoller$^{3,4}$, and G. Pupillo$^{3,4}$}

\affiliation{$^{1}$Department of Physics, University of
Massachusetts, Amherst, MA 01003, US}\affiliation{$^{2}$Institute
for Theoretical Physics III, University of Stuttgart,
Pfaffenwaldring 57, 70550 Stuttgart, Germany}

\affiliation{$^{3}$Institute for Theoretical Physics, University
of Innsbruck,  6020 Innsbruck, Austria}
\affiliation{$^{4}$Institute for Quantum Optics and Quantum
Information, 6020 Innsbruck, Austria}

\date{\today}

\begin{abstract}

We determine the phase-diagram of a one-dimensional system of
hard-core lattice bosons interacting via repulsive three-body
interactions by analytic methods and extensive quantum Monte-Carlo
simulations. Such three-body interactions can be derived from a
microscopic theory for polar molecules trapped in an optical
lattice. Depending on the strength of the interactions and the
particle density, we find superfluid and solid phases, the latter
appearing at an unconventional filling of the lattice and
displaying a coexistence of charge-density-wave and bond orders.
\end{abstract}

\maketitle
\section{Introduction}

Quantum many-body systems provide a wealth of fascinating
phenomena in condensed matter physics, including superfluidity in
liquid Helium, the fractional quantum Hall effect, as well as the
exotic electronic states in the pseudogap regime of cuprate
superconductors. While these quantum phases emerge from dominant
two-body interactions, with higher-order many-body interaction
terms providing only small corrections, an exciting recent avenue
of research in atomic and molecular physics is to engineer systems
where higher-order interactions dominate. In particular, it was
recently shown that this goal can be achieved for three-body
interactions using polar molecules \cite{BuchlerNature07}. In the
present work, we study the most fundamental model Hamiltonian
which displays three-body interactions in one dimension via
quantum Monte-Carlo simulations.
\\
\indent One-dimensional bosonic systems in the strongly correlated
regime have recently been realized with cold atomic gases:
examples are the superfluid/Mott-insulator quantum phase
transition for atoms trapped in optical
lattices~\cite{stoferle04},~\cite{fertig04}, and the cross-over
into the hard-core (Tonks-Girardeau) regime~\cite{kinoshita04}. A
characteristic feature of hard-core bosons in a lattice with
additional off-site two-body interactions is the appearance of
solid phases at half-filling $n=1/2$ with either  a charge-density
wave (CDW) or a bond-ordered (BOW) phase
\cite{schmitteckert},~\cite{moniensomma}.
\\ \indent
In contrast, here we study hard-core bosons with strong three-body
interactions. While the microscopic realization of the model with
polar molecules gives rise to next-nearest-neighbor two-body and
three-body interactions \cite{BuchlerNature07}, the dominant part
of the Hamiltonian is
\begin{equation}
 H = -J \sum_{i} \left[ b^{\dag}_{i} b^{}_{i+1} + b^{}_{i}b_{i+1}^{\dag} \right]
+ W \sum_{i} n_{i-1} n_{i} n_{i+1}.\label{eq:eqHam}
\end{equation}
The first term describes the standard kinetic energy with hopping
rate $J$, while the second term, accounts for the three-body
interaction with strength $W$; $n_{i} = b^{\dag}_{i} b_{i}$ is the
density operator with bosonic operators $b_{i}$ and $b^{\dag}_{i}$
satisfying the hard-core constraint. A similar model has recently
been studied in two-dimensions \cite{Schmidt08}.
\begin{figure}[htb!]
\begin{center}\includegraphics[width=0.8\columnwidth]{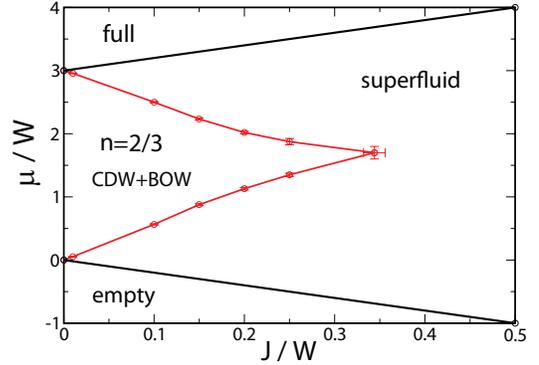}
\end{center}
\caption{(Color online) (QMC method: WA) Phase-diagram of
hard-core lattice bosons with dominant three-body interactions,
Eq.~\eqref{eq:eqHam}, in the grand-canonical ensemble, $\mu/W$ vs
$J/W$. The solid phase at filling $n=2/3$ is characterized by a
coexistence of CDW and BOW orders. }\label{fig:fig1}
\end{figure}
\\ \indent
We derive the complete quantum phase diagram of
Hamiltonian~\eqref{eq:eqHam} by means of extensive quantum
Monte-Carlo simulations. We find the existence of both superfluid
(SF) and solid phases, see Fig.~\ref{fig:fig1}. However, in
contrast to systems with two-body interactions, we show that the
solid phase appearing  at the unconventional filling $n=2/3$
exhibits both CDW and BOW orders.  While Luttinger liquid theory
predicts also instabilities towards a solid phase at $n=1/2$ and
$n=1/3$, we show here that the system remains superfluid even for
strong three-body interactions $W/J\gg 1$. Solid phases at filling
$n=1/2$ are found by adding {\em weak two-body} nearest-neighbor
(NN) and next-nearest-neighbor (NNN) corrections to
Eq.~\eqref{eq:eqHam}, $V_1 \sum_i n_i n_{i+1}$ and $V_2 \sum_i n_i
n_{i+2}$ respectively, as naturally realized with polar
molecules~\cite{BuchlerNature07}.
\begin{figure*}[ht]
\begin{center}
\includegraphics[width=0.9 \textwidth]{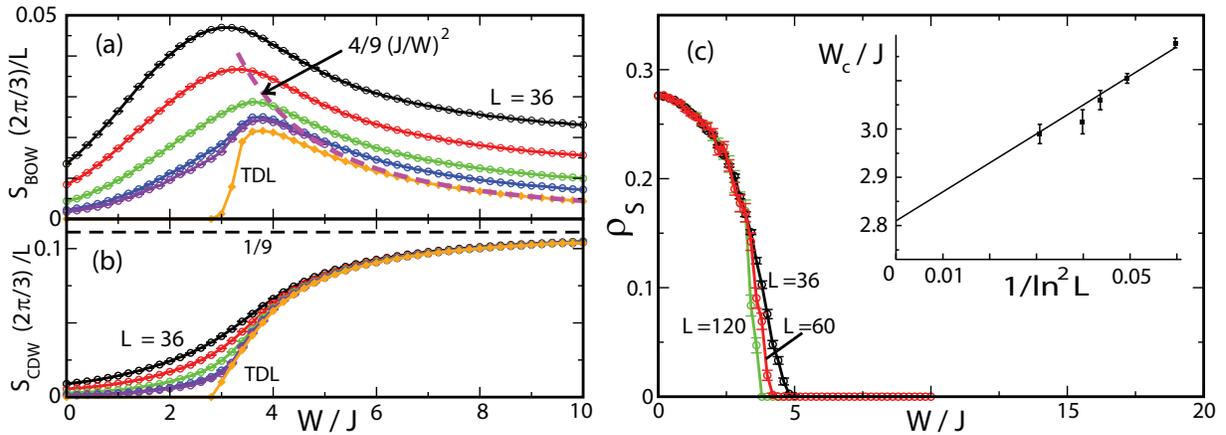}
\end{center}
\caption{(Color online) (QMC method: SSE) {\em Lattice filling
$n=2/3$.} (a) Bond-order structure factor $S_{\rm BOW}(2 \pi/3)/L$
vs $W/J$. Top to bottom: lattice sizes $L=36,60,120,240$ and 300;
the thermodynamic limit (TDL) is indicated. Dashed line: strong
coupling perturbative result $4/9 (J/W)^2$. (b) Density structure
factor $S_{\rm CDW}(2 \pi/3)/L$ vs $W/J$. Top to bottom: lattice
sizes $L=36,60,120,240$ and 300; the TDL is indicated. Dashed
line: strong coupling result $n^2/4=1/9$. (c) Superfluid density
$\rho_s$ as a function of $W/J$, for lattice sizes $L=36,60$ and
120. Inset: $W_{\rm c}/J$ as a function of $1/\ln^2 L$, results
from the WA. The line is a guide to the eye.}\label{fig:fig2}
\end{figure*}
\section{Results} The ground
state phase diagram of the model in Eq.~(\ref{eq:eqHam}) is
derived in the grandcanonical ensemble by varying the chemical
potential $\mu$ at different fixed values of $W/J$. We use two
different quantum Monte-Carlo (QMC) methods: (i) the stochastic
series expansion (SSE) algorithm with a generalized directed loop
update~\cite{sse2} after a decoupling of the Hamiltonian in
trimers for each tree-body interaction term, and (ii) a code based
on the Worm algorithm (WA) path integral approach~\cite{Worm1},
which allows efficient sampling of the many-body path winding
numbers in imaginary time and space directions. Although the SSE
method samples both BOW and CDW orders, its efficiency drops for
very large values of $W/J$. The WA does not suffer from this issue
and is especially useful in checking the limit $W/J\gg 1$. Results
obtained with the two methods are found to be consistent (see
below). The groundstate properties of the finite systems have been
probed using temperatures $T=0.6J/L$, with $L$ the number of
lattice sites, which was found sufficiently low.
\\ \indent The phase diagram is determined by two phases, see
Fig.~\ref{fig:fig1}: a superfluid Luttinger liquid (LL) phase with
algebraic correlations, surrounding a solid phase at filling
$n=2/3$, which appears for dominant three-body interactions $W/J
\gtrsim 3$. The solid phase is incompressible, giving rise to the
characteristic lobe structure in the $\mu$-$W$ phase diagram. This
incompressible phase is characterized by the structure factors
$S_{\rs CDW}$ for a charge density wave and $S_{\rs BDW}$ for a
bond order at the wave-vector $k = 2\pi /3$
\begin{eqnarray}
S_{\rs CDW}(k) &= & \frac{1}{L}\sum_{j,l} \exp\left[i
k\left(j-l\right)\right]
\langle n_{j} n_{l}\rangle, \label{eq:eqCDW}\\
S_{\rs BOW}(k) & = & \frac{1}{L} \sum_{j,l}  \exp\left[i
k\left(j-l\right)\right]
 \langle K_{j} K_{l} \rangle,\label{eq:eqBOW}
\end{eqnarray}
with the bond operators  $K_{l}= b_{l}^{\dag}b_{l+1}+ b_{l}
b^{\dag}_{l+1}$.
\begin{figure*}[ht]
\begin{center}
\includegraphics[width=0.9 \textwidth]{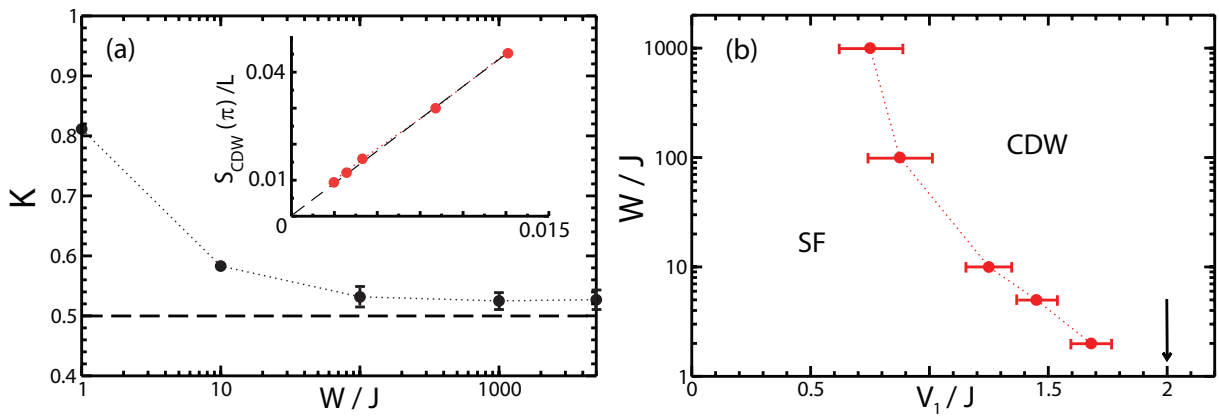}
\end{center}
\caption{{(Color online) (QMC method: WA) \em Lattice filling
$n=1/2$.} (a) Luttinger Liquid parameter $K$ as a function of
$W/J$ with critical value $K_{\rm c}=0.5$ (dashed line) of the
superfluid/solid transition. The inset shows the density structure
factor $S_{\rs CDW}(\pi)/L$ as a function of the inverse system
size $1/L$ for $W/J=1000$. (b) Phase diagram in the presence of an
additional nearest neighbor interaction $V_{1}$ in the plane $W/J$
vs $V_1/J$. The arrow indicates the point of the superfluid/solid
transition
at $W/J=0$.}\label{fig:fig3} \end{figure*}%
\\ \indent
The presence of CDW order can be easily understood in the limit
$J=0$, where the ground state at filling $n=2/3$ is threefold
degenerate. In fact, up to lattice translations, each groundstate
takes the form $ |\Omega\rangle = \prod_{k} b^{\dag}_{3 k}
b^{\dag}_{3 k+2} |0\rangle$, and exhibits CDW order with $ S_{\rs
CDW}(2 \pi/3) = n^2  L/4$. From the unconventional filling $n=2/3$
of the solid phase, it follows that once hopping of particles is
allowed the charge density order implies small hopping
correlations, since the position of a particle in the solid breaks
the symmetry of hopping to the left or hopping to the right. Using
standard perturbation theory in the hopping term, we find that
also a bond order wave appears for finite hopping with $S_{\rs
BOW}(2 \pi/3) = n^2 L J^2/W^2$.  Monte Carlo results directly
confirm the coexistence of the two orders. Analogous to the case
of the Hubbard-model with onsite two-body interactions, this
incompressible phase at filling $n=2/3$ can be reached by varying
the density, which corresponds to a mean-field transition, or by
keeping the density constant and by varying the strength of the
interactions $W/J$, see Fig.~\ref{fig:fig1}.  We found that the
superfluid density vanishes at the boundaries, indicating that
doping of the system happens simultaneously to the solid/SF
transition. In the following we will be mainly interested in
characterizing the constant-density transition at the tip of the
lobe.
\subsection{Filling $n=2/3$}
Our results for the order parameters $S_{\rs BOW}$ and $S_{\rs
CDW}$ at fixed density $n=2/3$ are shown in Fig.~\ref{fig:fig2}
for system sizes $L=36,60, 120, 240$ and 300 (top to bottom),
together with the extrapolated thermodynamic limit (TDL) behavior.
The latter has been obtained based on an observed linear scaling
of $S_{\rs BOW}$ and $S_{\rs CDW}$ in $1/L$ within the solid
phase, and  is found to be in perfect agreement with the
strong-coupling results for the order parameters valid for $W/J
\gg 1$ (dashed lines). In the thermodynamic limit, the two order
parameters are found to vanish simultaneously at a critical ratio
$W_c/J\approx 2.9$, corresponding to the solid/liquid transition
at the tip of the lobe.
\\ \indent
A more refined determination of the transition point can be
obtained using bosonization techniques.  In the weakly interacting
regime with  $W/J \ll 1$, the system can be mapped onto the
sine-Gordon model~\cite{BuchlerNature07}
\begin{equation}
H= \frac{\hbar v}{2} \int dx \left\{  \left[K \Pi^2 + \frac{1}{K}
\left(\partial_{x} \Phi\right)^2\right] +\lambda \cos (\gamma
\Phi) \right\},
\end{equation}
where $\Phi$ denotes the charging field and $\Pi$ the canonical
conjugate operator, while $K$ is the LL parameter, and $\gamma =
\sqrt{36 \pi}$ the periodicity of the sine-term.  Consequently,
the transition from the LL to the solid phase at the tip of the
lobe in Fig.~\ref{fig:fig1} appears at the critical value $K_{c} =
2/9$ and it belongs to the Kosterlitz-Thouless (KT) universality
class. For weak interactions $W/J \ll 1$ the behavior of the
Luttinger parameter is obtained  directly from  bosonization
techniques as $K=1- (2 \sqrt{3}/\pi) W/J $, however this value is
strongly renormalized close to the liquid/solid quantum phase
transition. We compute the LL parameter numerically as  $K=\hbar
\pi\sqrt{\rho_s\kappa /m}$ by QMC methods, with $\rho_s$ and
$\kappa$ the superfluid density and compressibility, respectively.
The latter are calculated from the statistics of winding numbers
$\left\langle W_{\alpha}^2 \right\rangle$ in imaginary time and
space directions. In particular, for a square system such that
$L_{\tau}\approx L_{x}=L$, with $L_{\tau}=\hbar v/T$ and $v=\sqrt{
\rho_s/\kappa  m }$ the sound velocity, $\rho_{s}=m L T\langle
W_x^2\rangle/\hbar^2 $~\cite{pollock}, and
$\kappa=\langle{W_{\tau}^2}\rangle /L T$, respectively, with
$m=\hbar^2/2 J$. We have performed simulations for $L=60, 90, 120,
150$ and 300. In order to precisely locate the critical point we
employ finite size scaling arguments following from the KT
renormalization group flow~\cite{KT}. Calling $W_c(L)$ the value
of $W$ for which $K(L)=K_c$, the finite size scaling of the
transition point is $W_c(L)-W_c\propto [\ln(L)]^{-2}$, with $W_c$
the transition point in the thermodynamic limit. The inset in
Fig.~\ref{fig:fig2}(c) shows the finite size scaling of $W_c(L)$,
which gives the transition point $W_c/J=(2.80\pm 0.15)$, in
agreement with the discussion above.
\subsection{Filling $n=1/2$}
\begin{figure}[t!]
\begin{center}
\includegraphics[width=0.85\columnwidth]{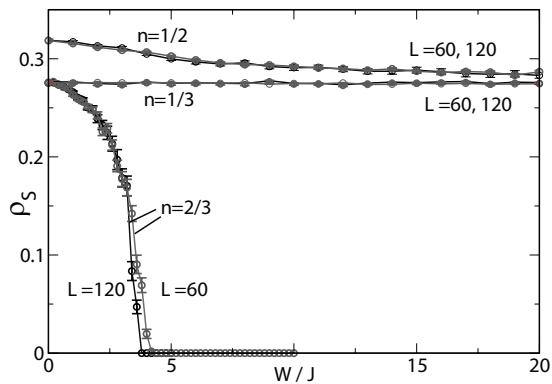}
\end{center}
\caption{(QMC method: SSE) Superfluid fraction $\rho_s$ as a
function of $W/J$ for fillings $n=1/3,1/2$ and $2/3$. Grey and
black lines correspond to system sizes $L=60$ and $120$,
respectively. Due to particle-hole symmetry, the superfluid
fractions for $n=1/3$ and $2/3$ are the same for $W/J =
0$.}\label{fig:fig4}
\end{figure}

In Ref.~\onlinecite{BuchlerNature07} it is argued that for lattice
filling $n=1/2$ a superfluid/solid transition should occur as a
function of $W/J$ at a critical LL parameter $K_{\rm c}=0.5$. This
instability is based on the observation that replacing the density
operator by its fluctuations around the mean value $n_{i}= n +
\delta n_{i}$, the three-body interaction gives rise to a nearest
neighbor and next-nearest neighbor interaction,
\begin{displaymath}
 W \sum_{i} n_{i-1}n_{i}n_{i+1} \sim W n \sum_{i }\left[  \delta n_{i-1} \delta n_{i+1} + 2   \delta
n_{i-1} \delta n_{i} \right].
\end{displaymath}
\\ \indent
In analogy to systems with two-body interactions, one would expect
a solid phase at half-filling $n=1/2$. However, we find that the
competition between the nearest-neighbor interaction of strength
$2 W n$, which drives an instability towards a CDW solid, and the
next-nearest neighbor interaction of strength $W n$, which drives
an instability towards a BOW solid, removes all instabilities
altogether, see below. The low-temperature phase is thus a
superfluid, independent of the magnitude of $W/J$. This behavior
is a special property of three-body interactions, and it is in
stark contrast to the two-body case, where a transition into the
solid phase has been always reported~\cite{schmitteckert}. In
Fig.~\ref{fig:fig3}(a) we show our results for  $K$ as a function
of $W/J$, obtained using the procedure described above for system
sizes $L=80,120,160,240$ and 320. For weak three-body interactions
$W/J\lesssim 1$, $K$ tends to the hard-core value $K=1$. We find
that for large interactions $W/J\gg 1$ the LL parameter saturates
at a value $K=(0.528 \pm 0.015)> K_c$ which is larger than the
critical value $K_c=0.5$, and thus no superfluid/solid transition
occurs at filling $n=1/2$. Consistently, the superfluid fraction
does not show any appreciable system-size dependence,
Fig.~\ref{fig:fig4}.
\\ \indent
The inset of Fig.~\ref{fig:fig3}(a) shows the  structure factor
$S_{\rs CDW}(\pi)$ as a function of the inverse system size $1/L$,
for the case of $W/J=1000$. The structure factor extrapolates to a
value consistent with zero in the termodynamic limit, consistent
with a superfluid in this limit of large three-body interactions.
It remains an open question, whether there is an analytical result
predicting the value of the Luttinger parameter for infinite
three-body interactions in analogy to hard-core bosons with $K=1$.
\subsection{Additional two-body interactions} We finally consider
the effects of additional short-ranged two-body repulsions, by
adding to the model in Eq.~(1) a term $V_1\sum_i n_i n_{i+1}+ V_2
\sum_i n_i n_{i+2}$ with nearest neighbor (NN) repulsion $V_1$,
and next-nearest-neighbor (NNN) repulsion $V_2$.  Such terms are
important as they constitute the leading correction to the
short-range three-body repulsions considered thus far, and can
induce new instabilities of the LL. In fact, it is well known that
a half-filled system of hard-core bosons with NN repulsions $V_1$
undergoes a superfluid/CDW KT transition at $V_1 = 2J$ to a
two-fold degenerate state with ordering pattern
$\circ\bullet\circ\bullet$ ($k=\pi$) \cite{Khono}.
\\ \indent
In Fig.~\ref{fig:fig3}(b) we show how this transition is modified
due to the presence of three-body interactions, by presenting the
low-energy phases as a function of $W/J$ and $V_1/J$. We find that
the presence of three-body interactions renormalizes considerably
the SF/CDW transition by shifting it to lower values of $V_1/J<2$.
In particular, for $W/J=1000$ we find $V_1^c/J=(0.75 \pm 0.15)$.
\\ \indent
The presence of NNN interactions $V_2 >0$ can drive the system
into a bond-ordered phase. In Ref.~\onlinecite{schmitteckert} it
is  shown that an ensemble of hard-core bosons with
NNN-interactions (at $V_1=0$) enters a $k=\pi$ bond-ordered phase
for $ V_2/J= (2.15\pm 0.10)$. For even larger $V_2/J>(2.66\pm
0.10)$, a four-fold degenerate $k=\pi/2$ CDW phase with ordering
pattern $\circ\bullet\bullet\circ$ becomes stabilized. We find
that the presence of three-body interactions significantly
renormalizes these transition points. For example, at $W/J=10$ the
SF/BOW transition occurs for $V_2/J=(0.9\pm0.1)$, and the BOW/CDW
transition at $V_2/J=(2.1\pm0.1)$. Three-body interactions thus
widen the extent of the BOW phase in this regime.
\subsection{Filling $n=1/3$}
We find that at filling $n=1/3$ the low-energy phase remains a
superfluid independently of the strength of the three-body
interactions. Consistently, the superfluid fraction $\rho_s$ in
this case is found to be essentially independent of $W/J$ in the
range $0<W/J<20$, as seen in Fig.~\ref{fig:fig4}. We computed the
value of the LL parameter for $W/J=1000$, and  found that $K=(0.89
\pm 0.01)> K_{\rm c}=2/9$, which confirms that three-body
interactions alone do not induce a transition in this region. This
finding contrasts the weak-coupling result
Ref.~\onlinecite{BuchlerNature07}, and exhibits the strong
breaking of particle-hole symmetry by finite three-body
interactions, as seen for $W\neq0$ from e.g. comparing $\rho_s$
for the two fillings $n=1/3$ and $n=2/3$ in Fig.~\ref{fig:fig4}.
Adding a NN interaction also does not induce any transition at
$n=1/3$. In fact, we find for $W/J=1000$ and $V_1/J=20$ that
$K=(0.48 \pm 0.01) > K_{\rm c}$.
\section{Conclusions}
We have determined the phase diagram of the fundamental model
Hamiltonian for hard-core bosons with dominant short-range
three-body interactions in one dimension. The latter can be
realized with polar molecules cooled to the electronic and
vibrational groundstate, which is well in the reach of current
experiments~\cite{Exp}. The one-dimensional nature of the problem
opens fascinating prospects for studying the dynamics of systems
with many-particle interactions by using powerful numerical
techniques such as e.g. tDMRG~\cite{Vidal}. Extensions to two
dimensions~\cite{Schmidt08} in various lattice geometries hold
promises in the search for exotic phases, such as e.g. topological
phases and spin liquids.

We acknowledge discussions with N.~V. Prokof'ev. This work was
supported by the ESF with  EuroQUAM, FWF, MURI, OLAQUI, and DARPA.
Allocation of CPU time on NIC J\"ulich and HLRS Stuttgart is
acknowledged.


\begin{thebibliography}{10}
\bibitem{BuchlerNature07}
H.~P. B{\"u}chler, A.~Micheli, and P.~Zoller,
\newblock Nature Physics {\bf 3}, 726 (2007).
\bibitem{stoferle04}
T.~St{\"o}ferle, Henning Moritz, Christian Schori, Michael
K{\"o}hl, and Tilman Esslinger,
\newblock Phys.\ Rev.\ Lett. {\bf 92}, 130403 (2004).
\bibitem{fertig04}
C.~D. Fertig, K. M. O'Hara, J. H. Huckans, S. L. Rolston, W. D.
Phillips, and J. V. Porto,
\newblock Phys.\ Rev.\ Lett. {\bf 94}, 120403 (2005).
\bibitem{kinoshita04}
T.~Kinoshita, T.~R. Wenger, and D.~S. Weiss,
\newblock Science {\bf 305}, 1125 (2004); B.~Paredes, A. Widera, V. Murg,
O.Mandel, S. F\"{o}lling, I. Cirac, G. Shlyapnikov, T. H\"{a}nsch,
and I. Bloch,
\newblock Nature {\bf 429}, 277 (2004).
\bibitem{schmitteckert}
P.~Schmitteckert and R.~Werner,
\newblock Phys. Rev. B {\bf 69}, 195115 (2004).
\bibitem{moniensomma}
T. D. K\"uhner and H. Monien
\newblock Phys. Rev. B {\bf 58}, R14741 (1998);
R. D. Somma and A.A. Aligia,
\newblock Phys. Rev. B {\bf 64}, 024410 (2001).
\bibitem{Schmidt08}
K.~P. Schmidt, J.~Dorier, and A.~M. {L\"a}uchli,
\newblock Phys. Rev. Lett. {\bf 101}, 150405 (2008).
\bibitem{sse2}
A.~W. Sandvik,
\newblock Phys. Rev. B {\bf 59}, R14157 (1999);
O.~F. Sylju{\aa}sen and A.~W. Sandvik,
\newblock Phys. Rev. E {\bf 66}, 046701 (2002).
\bibitem{Worm1}
N.~V. Prokof'ev, B.~V. Svistunov, and I.~S. Tupitsyn,
\newblock Phys. Lett. A {\bf 238}, 253 (1998); JETP {\bf 87}, 310 (1998).
\bibitem{pollock}
E.~L. Pollock and D.~M. Ceperley,
\newblock Phys. Rev. B {\bf 36}, 8343 (1987).
\bibitem{KT} V.L.~Berezinskii, JETP {\bf 34}, 610 (1972); J.~M. Kosterlitz and D.J. Thouless,
\newblock J.Phys. C {\bf 6}, 1181 (1973).
\bibitem{Khono} M. Kohno and M. Takahashi, Phys. Rev. B {\bf 56},
3212 (1997).
\bibitem{Exp} S. Ospelkaus, A. Pe'er, K.-K. Ni, J. J. Zirbel, B. Neyenhuis,
S. Kotochigova, P. S. Julienne, J. Ye, and D. S. Jin, Nature
Physics {bf 4}, 622 (2008); J. G. Danzl, E. Haller, M. Gustavsson,
M. J. Mark, R. Hart, N. Bouloufa, O. Dulieu, H. Ritsch, and H-C.
N\"{a}gerl, Science {\bf 321}, 1062 (2008); J. Deiglmayr, A.
Grochola, M. Repp, K. M\"{o}rtlbauer, C. Gl\"{u}ck, J. Lange, O.
Dulieu, R. Wester, and M. Weidem\"{u}ller (2008).
\bibitem{Vidal}
G.~Vidal
\newblock Phys. Rev. Lett. {\bf 91}, 147902 (2003).


\end{thebibliography}
\end{document}